\documentclass[a4paper, 11pt]{article}

\usepackage{jinstpub}

\usepackage{lineno, hyperref}
\modulolinenumbers[5]
\usepackage{amsmath}
\usepackage{siunitx}
\usepackage{booktabs}
\usepackage{multirow}
\usepackage{subcaption}
\usepackage{color, soul}

\title{\boldmath Characterisation of resistive MPGDs with 2D readout}

\author[a]{L. Scharenberg,\note{Corresponding author.}}

\author[a]{F. Brunbauer,}
\author[a]{H. Danielson,}
\author[b,c]{Z. Fang}
\author[a,d]{K. J. Fl\"{o}thner,}
\author[e]{F. Garcia,}
\author[a,f,g]{D. Janssens,}
\author[a,h]{M. Lisowska,}
\author[b,c]{J. Liu,}
\author[b,c]{Y. Lyu}
\author[a]{B. Mehl,}
\author[a,i]{H. Muller,}
\author[a]{R. de Oliveira,}
\author[a]{E. Oliveri,}
\author[a,j]{G. Orlandini,}
\author[k,a]{D. Pfeiffer,}
\author[a]{O. Pizzirusso,}
\author[a]{L. Ropelewski,}
\author[k,a]{J. Samarati,}
\author[b,c]{M. Shao,}
\author[a]{A. Teixeira,}
\author[a]{M. Van Stenis,}
\author[a]{R. Veenhof,}
\author[b,c]{Z. Zhang,}
\author[b,c]{and Y. Zhou}

\affiliation[a]{European Organization for Nuclear Research (CERN), 1211 Geneva 23, Switzerland}
\affiliation[b]{State Key Laboratory of Particle Detection and Electronics, University of Science and Technology of China, Hefei 230026, China}
\affiliation[c]{Department of Modern Physics, University of Science and Technology of China, Hefei 230026, China}
\affiliation[d]{Helmholtz-Institut f\"{u}r Strahlen- und Kernphysik, University of Bonn, Nu\ss{}allee 14-16, 53115 Bonn, Germany}
\affiliation[e]{Helsinki Institute of Physics, P.O. Box 64, FI-00014 University of Helsinki, Finland}
\affiliation[f]{Inter-University Institute For High Energies, Pleinlaan 2, 1050 Brussels, Belgium}
\affiliation[g]{Vrije Universiteit Brussel, Pleinlaan 2, 1050 Brussels, Belgium}
\affiliation[h]{Universit\'{e} Paris-Saclay, F-91191 Gif-sur-Yvette, France}
\affiliation[i]{Physikalisches Institut, University of Bonn, Nu{\ss}allee 12, 53115 Bonn, Germany}
\affiliation[j]{Friedrich-Alexander-Universit{\"a}t Erlangen-N{\"u}rnberg, Schlo\ss{}platz 4, 91054 Erlangen, Germany}
\affiliation[k]{European Spallation Source ERIC (ESS), Box 176, SE-221 00 Lund, Sweden}

\emailAdd{lucian.scharenberg@cern.ch}

\abstract{
    Micro-Pattern Gaseous Detectors (MPGDs) with resistive anode planes provide intrinsic discharge robustness while maintaining good spatial and time resolutions.
    Typically read out with 1D strips or pad structures, here the characterisation results of resistive anode plane MPGDs with 2D strip readout are presented.
    A \textmu{}RWELL prototype is investigated in view of its use as a reference tracking detector in a future gaseous beam telescope.
    A MicroMegas prototype with a fine-pitch mesh (730 line-pairs-per-inch) is investigated, both for comparison and to profit from the better field uniformity and thus the ability to operate the detector more stable at high gains.
    Furthermore, the measurements are another application of the RD51 VMM3a/SRS electronics.
}


\begin{document}

\maketitle
\flushbottom


\section{Introduction}
In recent years, various experiments and detector R\&D lines started the use of Micro-Pattern Gaseous Detectors (MPGDs) with resistive elements, mainly because of their discharge robustness \cite{nsw,t2k,rhum,urwell-lhcb}.
Additionally, parameters such as the signal induction, i.e. the spread of the charge over a given number of readout channels, and rate-capability can be tuned to the desired values of the experiment by adjusting the design of resistive elements.
This is most prominently employed in MPGDs with a single amplification stage, such as MicroMegas (MM) \cite{micromegas} and \textmu{}RWELL \cite{urwell}.

In this paper, the results of a characterisation of these two technologies are shown.
The studies are conducted in view of various R\&D aspects.
At first, there is the technology of the readout anode, a 2D strip structure underneath a single layer of resistive material.
So far, most resistive MPGDs employ either 1D strip structures underneath the resistive layer, 1D resistive strips or resistive pads.
The second aspect is specific to the \textmu{}RWELL detector that has been investigated.
It is a prototype of what is supposed to be used as reference detectors for particle trajectory reconstruction in a new beam telescope of the DRD1 collaboration \cite{drd1-website,drd1-proposal} --- the preceding RD51 collaboration \cite{website_rd51,article_rd51} provided for its joined test beam campaigns at the H4 beam line of CERN's SPS two beam telescopes, one based on MicroMegas and one based on triple-GEM detectors \cite{article_vmm-mpgd}.
While a previous prototype with 2D X-Y strip readout showed good performance \cite{yi}, it showed an unequal signal sharing between the top strip layer and the perpendicular bottom strip layer.
For the new prototype which was investigated for the here presented studies, this imbalance has been addressed.
The third point is specific to the investigated MicroMegas detector.
The particular detector is equipped with a fine mesh in the amplification stage with 730 line-pairs-per-inch (LPI), corresponding to a mesh cell pitch of $\SI{35}{\micro m}$.
The motivation for investigating a mesh structure like this is a more uniform electric field and thus the possibility to operate the detector at higher gains.
While the stability measurements are part of a separate study --- here only the results on the spatial resolution, time resolution and charge behaviour are presented --- it can be noted that the MicroMegas detector could be operated at gains more than twice as high as the \textmu{}RWELL detector.
Another more general aspect is the application of the new VMM3a/SRS front-end electronics \cite{article_vmm-twepp} of the RD51 collaboration to a wider range of detectors.

\section{Experimental methods}

\subsection{Devices under test}
\begin{figure}[t!]
    \centering
    \begin{subfigure}{\columnwidth}
        \centering
        \includegraphics[width = \columnwidth]{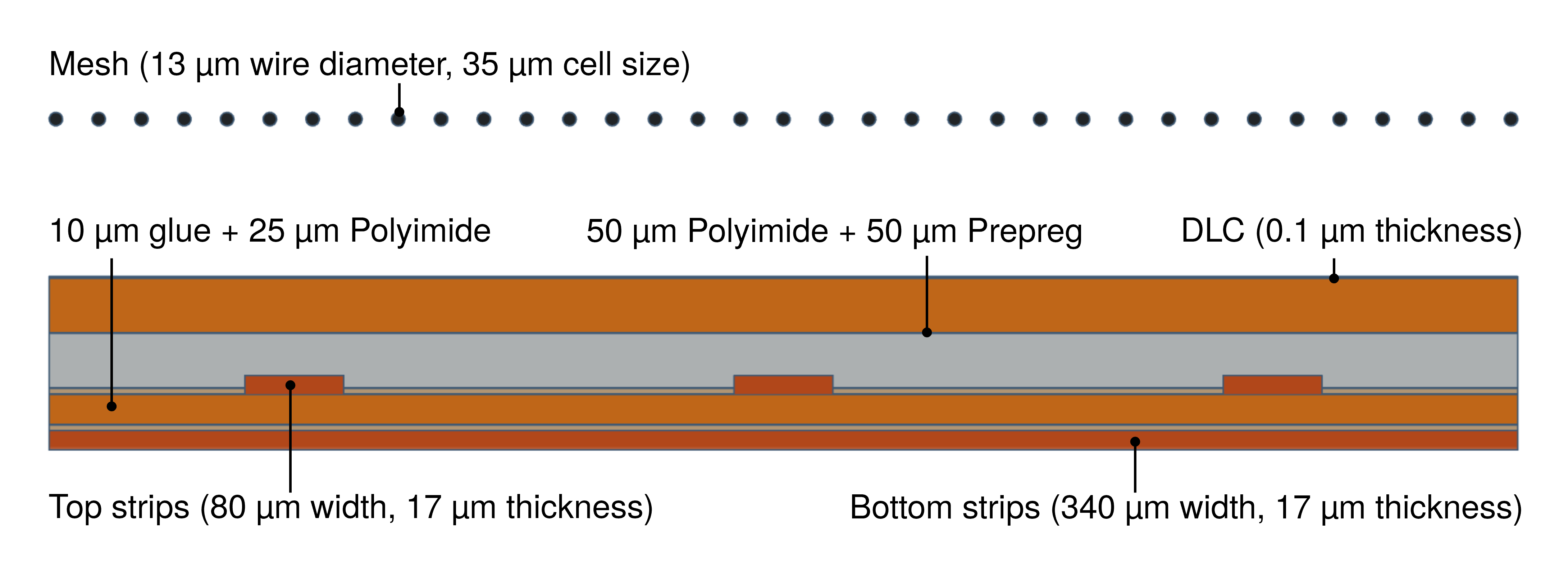}
        \caption{MicroMegas}
        \label{fig:detector-sketch-MM}
    \end{subfigure}
    \begin{subfigure}{\columnwidth}
        \centering
        \includegraphics[width = \columnwidth]{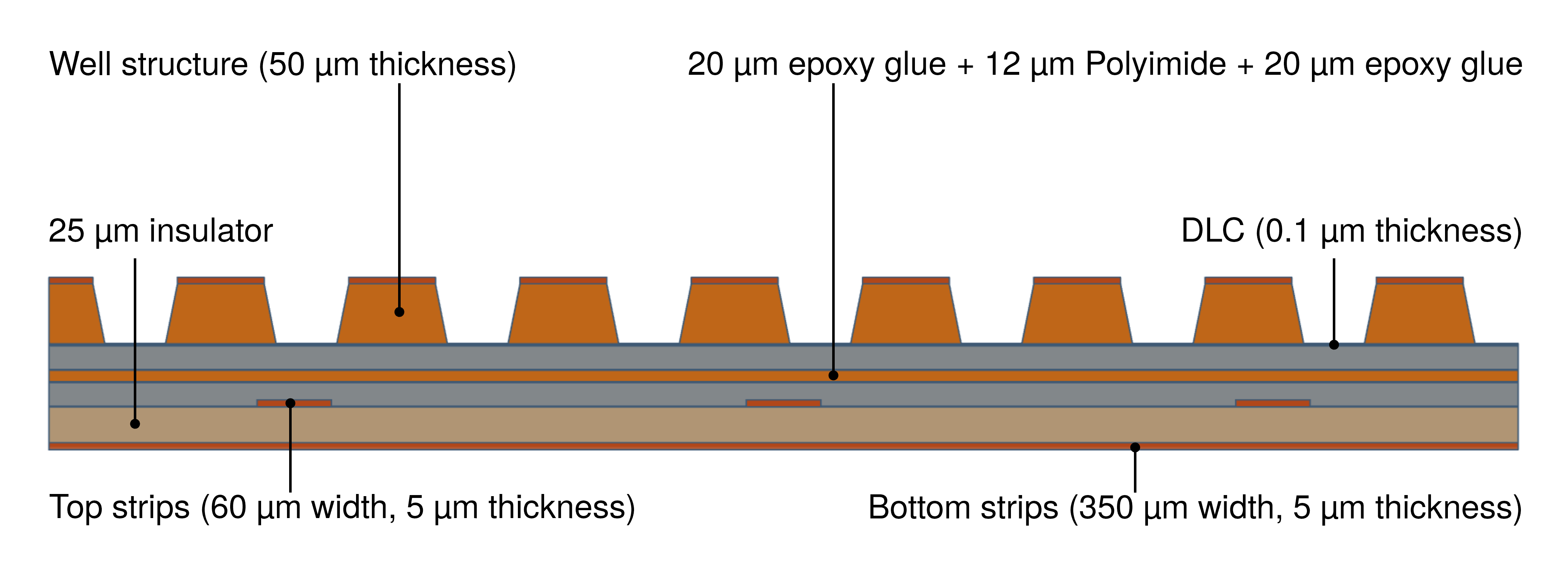}
        \caption{\textmu{}RWELL}
        \label{fig:detector-sketch-U}
    \end{subfigure}
    \caption{Sketched cross-section of the MicroMegas (a) and the \textmu{}RWELL (b) amplification stages, as used in the here presented measurements.}
    \label{fig:detector-sketch}
\end{figure}

Both detectors (sketched in Fig.~\ref{fig:detector-sketch}) have an active area of $\num{10}\times\SI{10}{cm^2}$ with a 2D X-Y-strip readout with 256 strips in each direction.
The strip pitch is $\SI{400}{\micro m}$.
In both detectors, the anode is a layer of Diamond-Like Carbon (DLC) with similar surface resistivities --- $\SI{40}{M\ohm/sq}$ for the \textmu{}RWELL and $\SI{37}{M\ohm/sq}$ for the MicroMegas.
The drift region of each detector had a width of $\SI{3}{mm}$.
Although both detectors were operated with a negative high voltage on the cathode, the MicroMegas detector was operated with a grounded mesh and a positive high voltage on the DLC anode, while the \textmu{}RWELL was operated with a negative high voltage at the top layer of the Well structure and a grounded DLC anode.

When performing gain scans, the drift field was kept constant at around $\SI{2.4}{kV/cm}$ for the \textmu{}RWELL, which corresponds to $\SI{724}{V}$ difference between drift cathode and Well, and $\SI{0.9}{kV/cm}$ for the MicroMegas detector, which corresponds to $\SI{275}{V}$ between cathode and mesh.
The amplification voltages were varied from $\num{460}$ to $\SI{570}{V}$ for the \textmu{}RWELL and $\num{580}$ to $\SI{700}{V}$ for the MicroMegas --- at higher voltages, the detectors started to show instabilities in operation due to discharges leading to high voltage trips.
When performing drift scans, the amplification voltage was kept constant at $\SI{680}{V}$ for the MicroMegas, which corresponds to an effective gain of around $\num{40000}$ at $\SI{0.9}{kV/cm}$ drift field, and at $\SI{540}{V}$ for the \textmu{}RWELL, which corresponds to a gain of around $\num{15000}$ at $\SI{2.4}{kV/cm}$ drift field.

\subsection{Beam telescope and readout electronics}
Both detectors have been characterised as Devices Under Test (DUTs) with the RD51 VMM3a/SRS beam telescope \cite{article_vmm-mpgd}.
It consists of three COMPASS-like triple-GEM detectors \cite{compass-gem} with an active area of $\num{10}\times\SI{10}{cm^2}$ almost equally spaced with a total lever arm of around $\SI{1}{m}$ to provide the position information.
In addition, it contains three scintillators with Photo-Multiplier Tubes (PMTs) connected to a NIM coincidence unit as a time reference.
All the MPGDs in the telescope have been operated with the same gas mixture of Ar/CO\textsubscript{2} (70/30\,\%).

For the readout of the detectors, including the output of the NIM coincidence unit, the ATLAS/BNL VMM3a front-end ASIC \cite{vmm} in its integration \cite{michael} of the RD51 Scalable Readout System (SRS) \cite{srs} has been used.
It provides the acquired charge per channel (10-bit ADC, effectively 7-bit) in a continuous, multi-channel self-triggered readout mode with around $\SI{1}{ns}$ time resolution and a MHz rate-capability.
The adjustable analogue front-end parameters of the VMM3a have been set to $\SI{200}{ns}$ peaking time, $\SI{9}{mV/fC}$ electronics gain and around $\num{10000}$ electrons threshold per channel for both the DUTs and the reference detectors.
The threshold level might seem high, but it should be considered that the VMM3a is operated in its self-triggered readout, meaning that each signal above the Threshold Level (THL) will be processed and become part of the data stream, i.e. a THL that is set too low will results in a lot of external pick-up noise being part of the raw data.
Furthermore, in terms of the dynamic range of the VMM3a, the THL is still a low value of less than $\SI{2}{\percent}$ per channel.

\section{Charge behaviour}

\begin{figure}[t!]
    \centering
    \begin{subfigure}{0.4\columnwidth}
        \centering
        \includegraphics[width = \columnwidth]{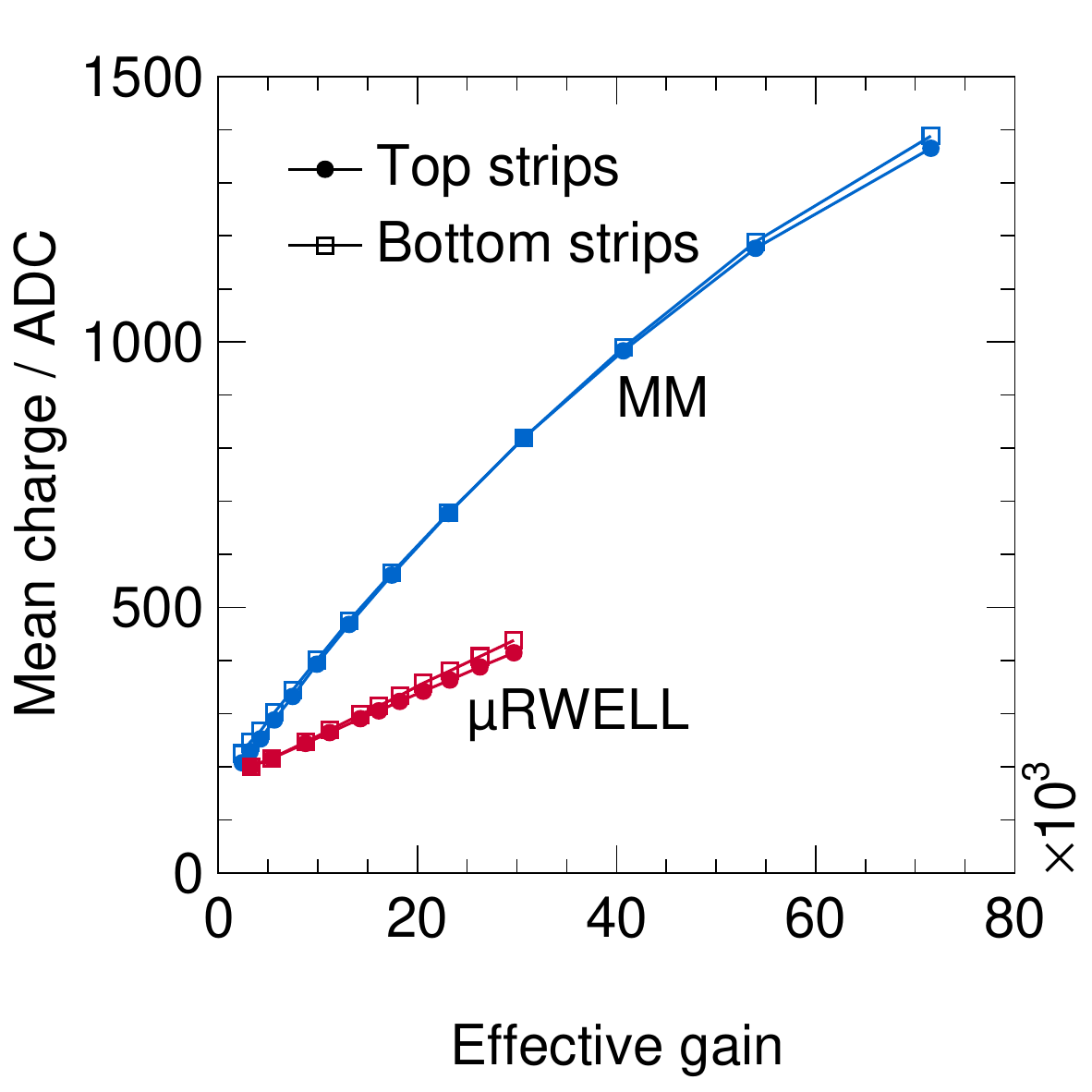}
        \caption{Gain dependence}
        \label{fig:charge-sharing}
    \end{subfigure}
    \hspace{0.1\columnwidth}
    \begin{subfigure}{0.4\columnwidth}
        \centering
        \includegraphics[width = \columnwidth]{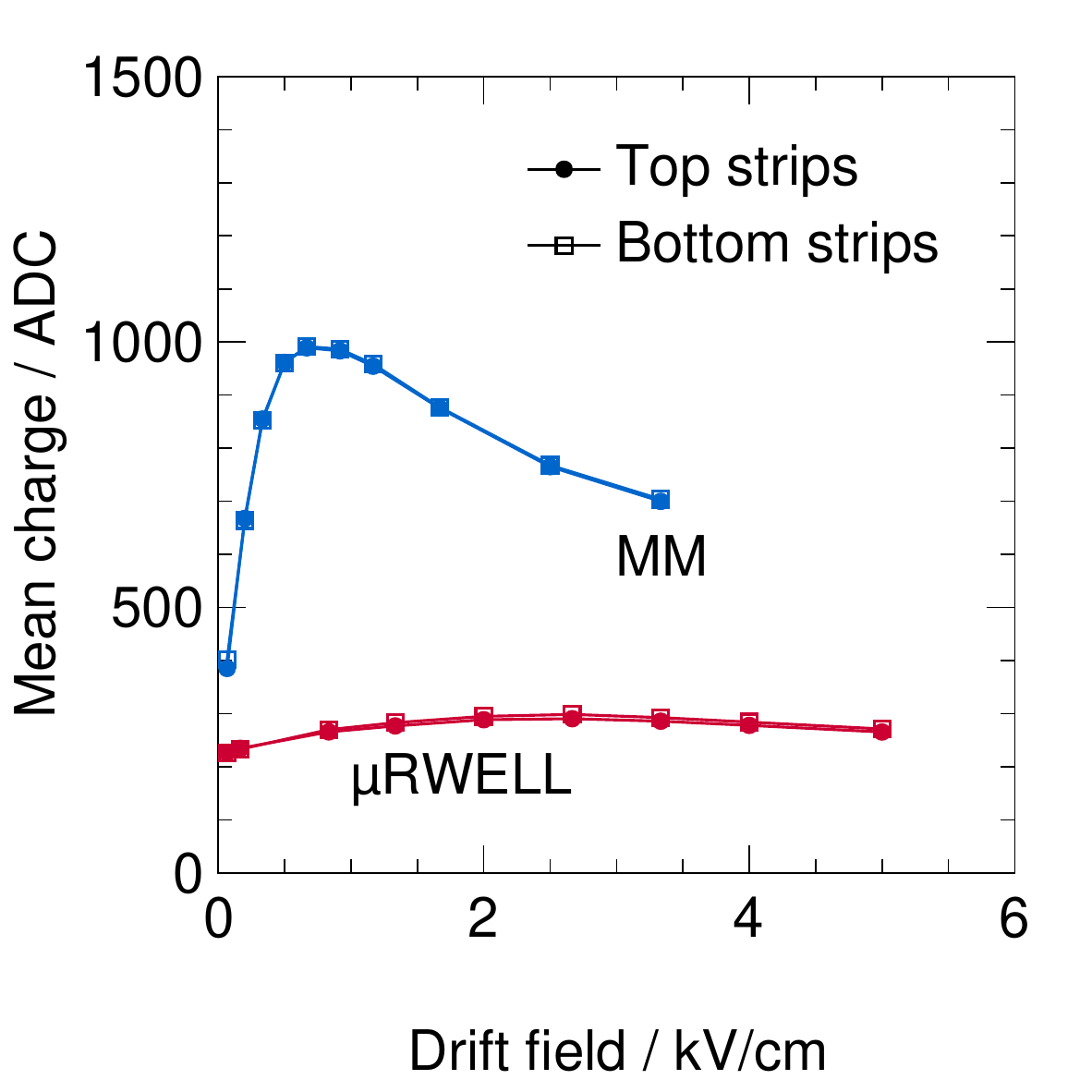}
        \caption{Drift field dependence}
        \label{fig:charge-collection}
    \end{subfigure}
    \caption{Average measured charge, i.e. the calculated mean of the energy loss distribution after amplification, depending (a) on the effective gain of the two investigated detectors or (b) on the drift field, in each read out detector plane.}
\end{figure}

With the previous \textmu{}RWELL prototype showing an imbalance of the charge collection between the top and the bottom readout strips \cite{yi}, the first point of investigation is the charge sharing and collection behaviour of the detectors.
For this, the charge distribution (Landau distribution) is generated from all recorded interactions that can be assigned to a reconstructed particle trajectory from the beam telescope.
Then the mean value of this measured distribution is taken and plotted depending on the effective gain of the detector (Fig.~\ref{fig:charge-sharing}).
It can be seen that both strip planes collect almost the same amount of charge, resulting in a charge-sharing ratio between the top and bottom strips of each detectors close to one.
Another observation is that the mean charge measured with the Landau distribution increases linearly with the gain for the \textmu{}RWELL detector, while for the MicroMegas detector, the increase starts to flatten out at higher gains.
This is due to the ADC saturation of individual readout channels which at higher gains is more likely to occur.

Another observation is that the total collected charge for the \textmu{}RWELL is significantly less than for the MicroMegas detector, despite the same effective gain.
This is due to the signal induction from the anode to the readout strips.
While the effective detector gain was determined from the current measured on the resistive anode layer, the front-end electronics only measures the induced current in the capacitively coupled readout strips.
This is important to note, as the gain at which a discharge might occur in the detector is the one determined via the signal current on the anode plane.

\begin{figure}[t!]
    \centering
    \begin{subfigure}{0.4\columnwidth}
        \centering
        \includegraphics[width = \columnwidth]{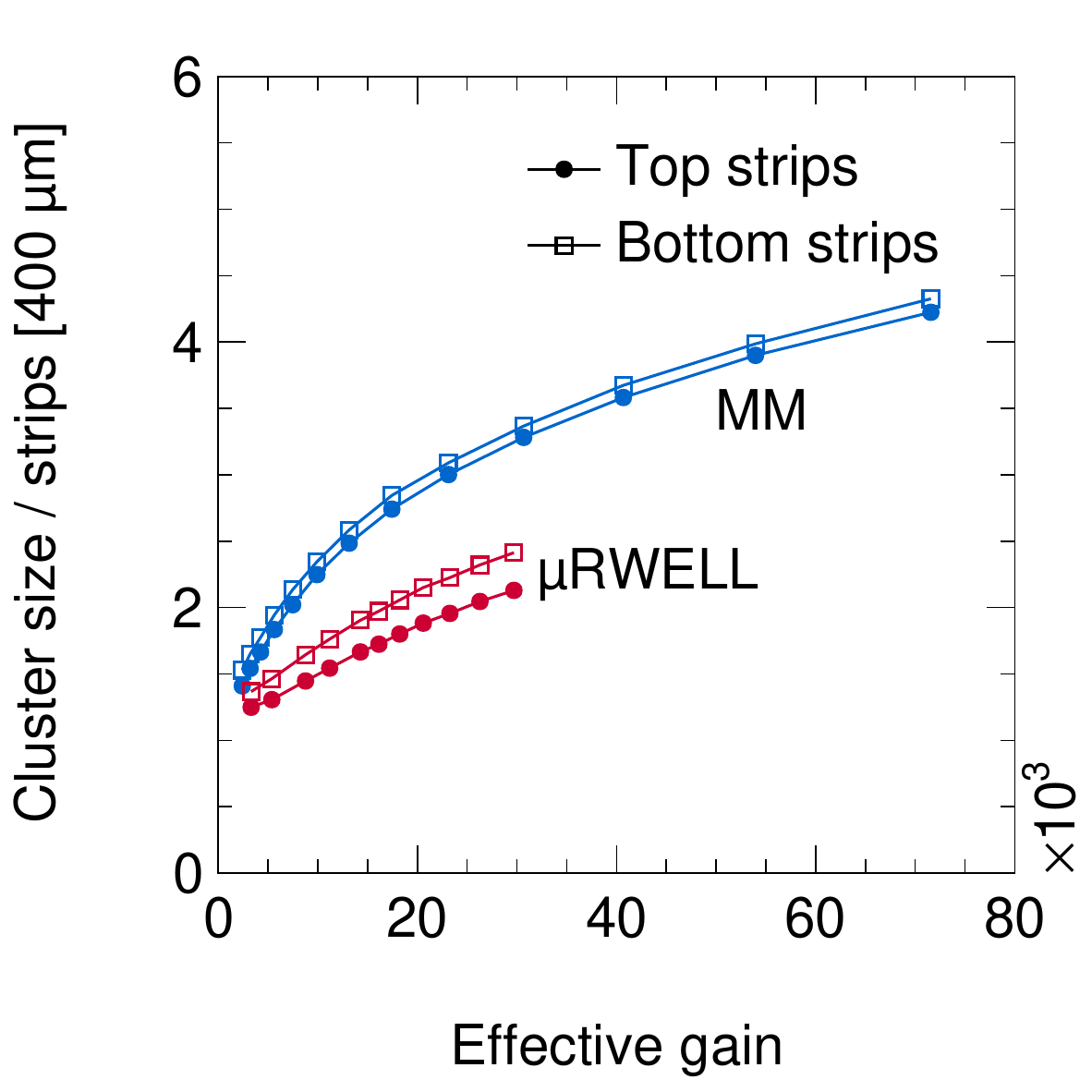}
        \caption{Average cluster size}
        \label{fig:cluster-size}
    \end{subfigure}
    \hspace{0.1\columnwidth}
    \begin{subfigure}{0.4\columnwidth}
        \centering
        \includegraphics[width = \columnwidth]{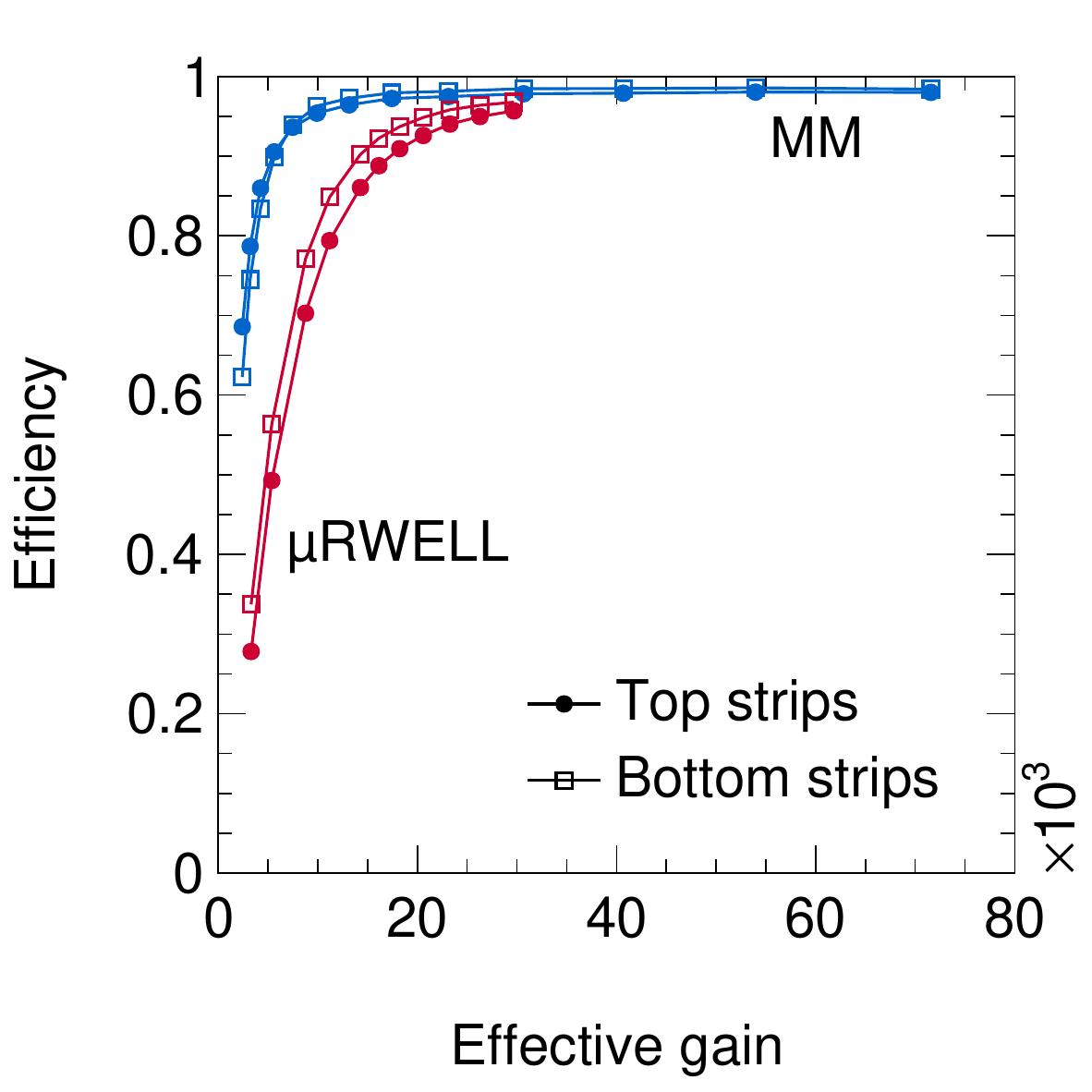}
        \caption{Detector efficiency}
        \label{fig:efficiency}
    \end{subfigure}
    \caption{In (a), the mean of the cluster size distribution (distribution of the number of channels above the THL for each recorded interaction) is shown, depending on the effective gain of the detectors.
    In (b), the detector efficiency as defined in Eq.~(\ref{eq:efficiency}) is shown, depending on the effective gain.}
\end{figure}

This explanation is strengthened by the behaviour of the cluster size (number of channels above the THL), with the mean cluster size being plotted against the effective detector gain (Fig.~\ref{fig:cluster-size}).
It can be seen that the cluster size scales accordingly, with the bottom strips of the \textmu{}RWELL collecting slightly more charge, also having a slightly larger cluster size.
And an almost doubled cluster size for the MicroMegas detector at similar gains, leading to almost twice the amount of measured charge.
This behaviour is also reflected in the detector efficiency (Fig.~\ref{fig:efficiency}), which is defined as
\begin{align}
    \epsilon_\mathrm{det} = \frac{N_\mathrm{det}}{N_\mathrm{tracks}} \ .
    \label{eq:efficiency}
\end{align}
Here, $N_\mathrm{tracks}$ is the number of particle interactions and trajectories reconstructed with the reference tracking detectors and $N_\mathrm{det}$ is the number of particle interactions that have been recorded in the DUT and that can be assigned to an existing trajectory.
Due to the difference in signal induction between the investigated \textmu{}RWELL and MicroMegas detectors, reflected by the smaller cluster size and measured signal amplitude for the \textmu{}RWELL, higher detector gains are needed to reach comparable cluster sizes and induced signal charges and thus comparable efficiencies.

In addition to the dependence on the effective gain of the detector, also the dependence of the measured charge depending on the electric drift field was studied, representing the charge collection behaviour by the amplification stage.
The results are shown in Fig.~\ref{fig:charge-collection}.
Both detectors show the expected characteristic behaviour, with the charge collection of the \textmu{}RWELL having a much broader peak at higher electric fields (e.g.\ as shown in \cite{urwell}), while the MicroMegas shows a more well define peak at lower electric fields (e.g.\ as shown in \cite{thesis_jona}).

\section{Spatial resolution}

\subsection{Basic results}

The spatial resolution of the DUTs is extracted from the width of the residual distributions that are generated from the difference $\Delta x = x_\mathrm{ref} - x'$ between the reference particle position $x_\mathrm{ref}$ that is provided by the reconstructed trajectory from the reference tracking detectors and the interaction point $x'$ reconstructed in the DUTs.
To determine the width, two overlapping Gaussian functions are fitted to the residual distribution, with the mean value being identical, but a weight factor $w$ to account for the different scales and different standard deviations $\sigma$ to account for the core and the tails of the residual distribution.
The final width is thus defined as
\begin{align}
    \sigma_{\Delta x}^2 = w\sigma_\mathrm{core}^2 + (1 - w)\sigma_\mathrm{tail}^2 \ .
\end{align}
The spatial resolution is then obtained by quadratically subtracting the contribution from the uncertainty on the track reconstruction, as described in \cite{thesis_horvat,thesis_jona,mythesis}.

\begin{figure}[t!]
    \centering
    \begin{subfigure}{0.4\columnwidth}
        \centering
        \includegraphics[width = \columnwidth]{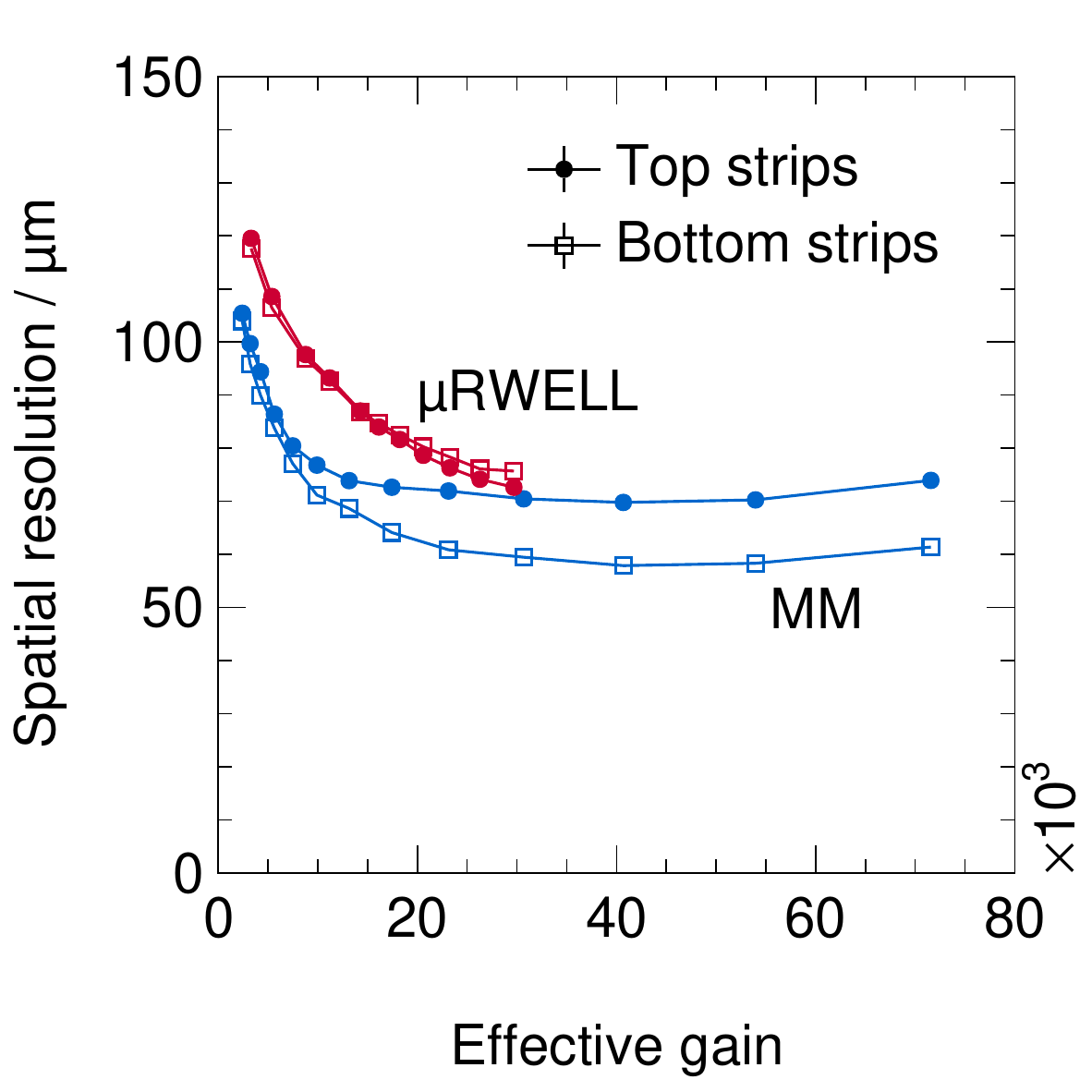}
        \caption{Gain dependence}
        \label{fig:spatial-resolution-cog}
    \end{subfigure}
    \hspace{0.1\columnwidth}
    \begin{subfigure}{0.4\columnwidth}
        \centering
        \includegraphics[width = \columnwidth]{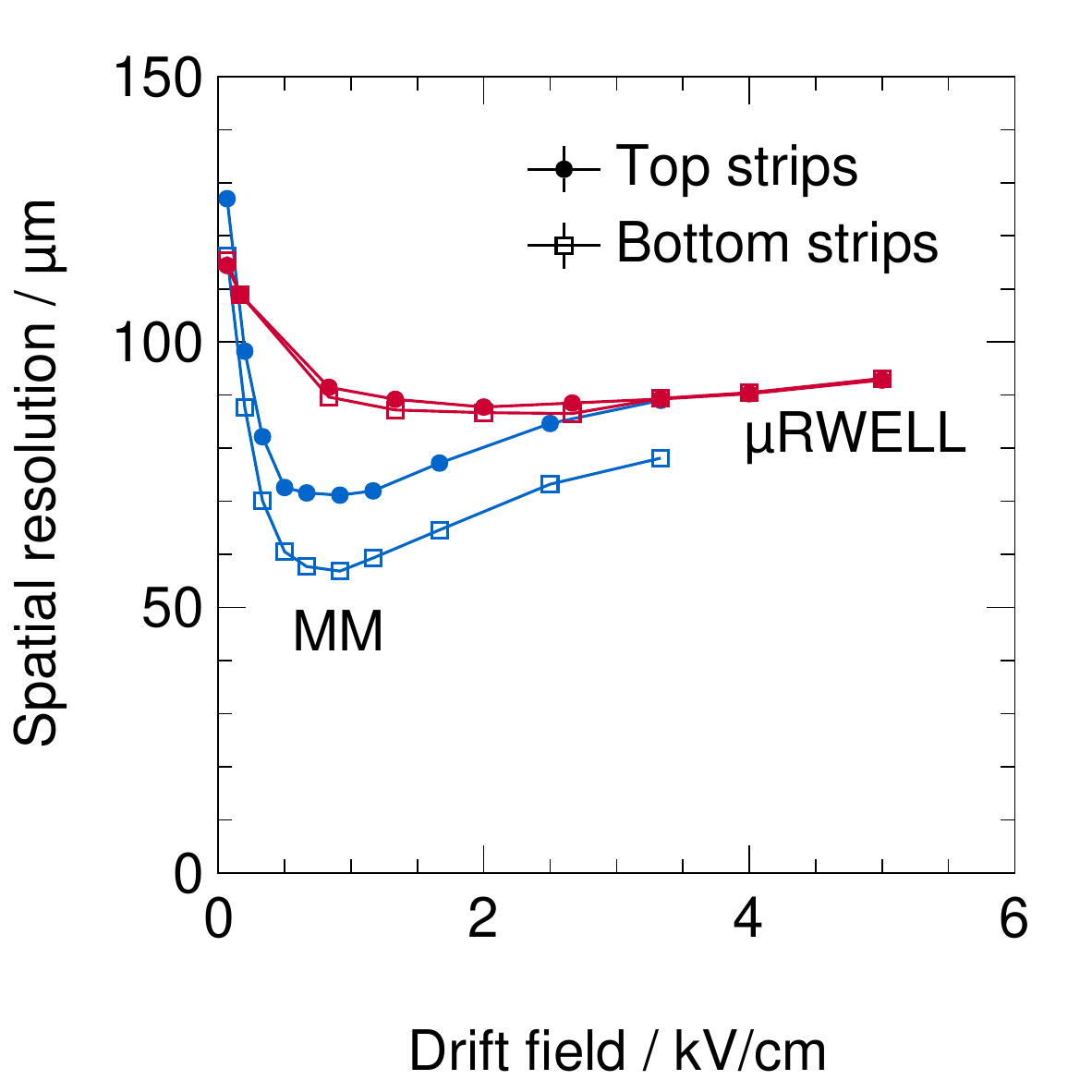}
        \caption{Drift field dependence}
        \label{fig:spatial-resolution-drift-cog}
    \end{subfigure}
    \caption{Dependence of the spatial resolution of the two DUTs on (a) their effective detector gain and on (b) the electric drift field, determined for both readout planes individually.}
\end{figure}

Using the Centre-Of-Gravity (COG) to determine the position of the particle interaction within the detector, the gain dependence of the spatial resolution as shown in Fig.~\ref{fig:spatial-resolution-cog} is obtained.
For both detectors, the spatial resolution improves with increased detector gain.
Taking the cluster size behaviour (Fig.~\ref{fig:cluster-size}) into account, this is related to more charge information being available and distributed over more than a single readout strip.
This also explains the better spatial resolution of the MicroMegas compared to the \textmu{}RWELL, as it has a larger charge collection and cluster size.
A similar dependence can be observed when plotting the drift behaviour (Fig.~\ref{fig:spatial-resolution-drift-cog}).
The spatial resolution is inversely proportional to the charge collection depending on the electric drift field (Fig.~\ref{fig:charge-collection}), i.e. the more charge collected, the better the spatial resolution.

At very high gains (Fig.~\ref{fig:spatial-resolution-cog}), the spatial resolution starts to decrease for the MicroMegas detector.
This is due to the saturated readout channel that also caused the flattening of the measured charge-gain-dependence (Fig.~\ref{fig:charge-sharing}).
The position reconstruction by COG loses accuracy when the relative amount of charge is not correctly represented anymore because the corresponding readout channel is in saturation.

\begin{figure}[t!]
    \centering
    \begin{subfigure}{0.4\columnwidth}
        \centering
        \includegraphics[width = \columnwidth]{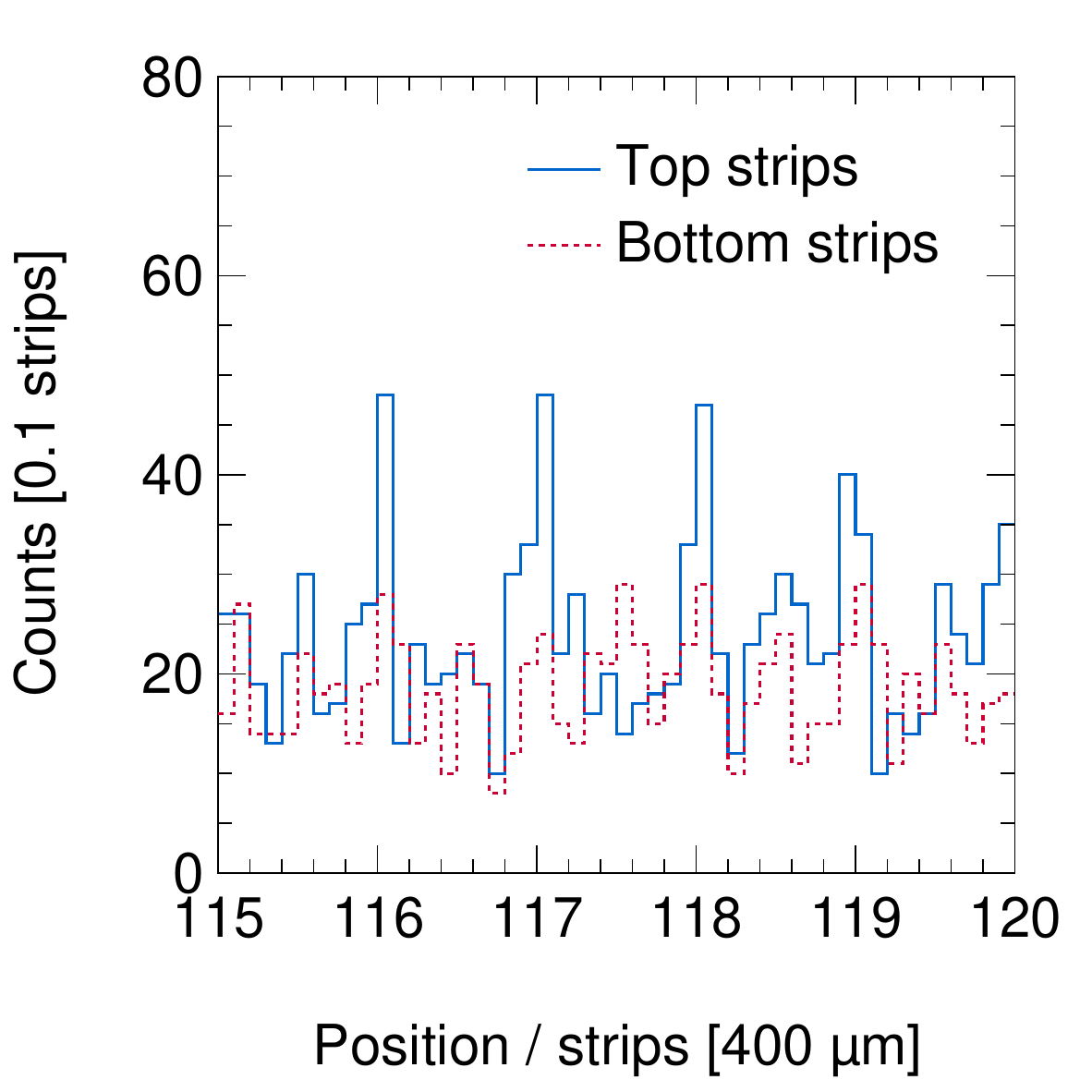}
        \caption{High gain, approximately $\num{40000}$}
        \label{fig:readout-modulation-high-gain}
    \end{subfigure}
    \hspace{0.1\columnwidth}
    \begin{subfigure}{0.4\columnwidth}
        \centering
        \includegraphics[width = \columnwidth]{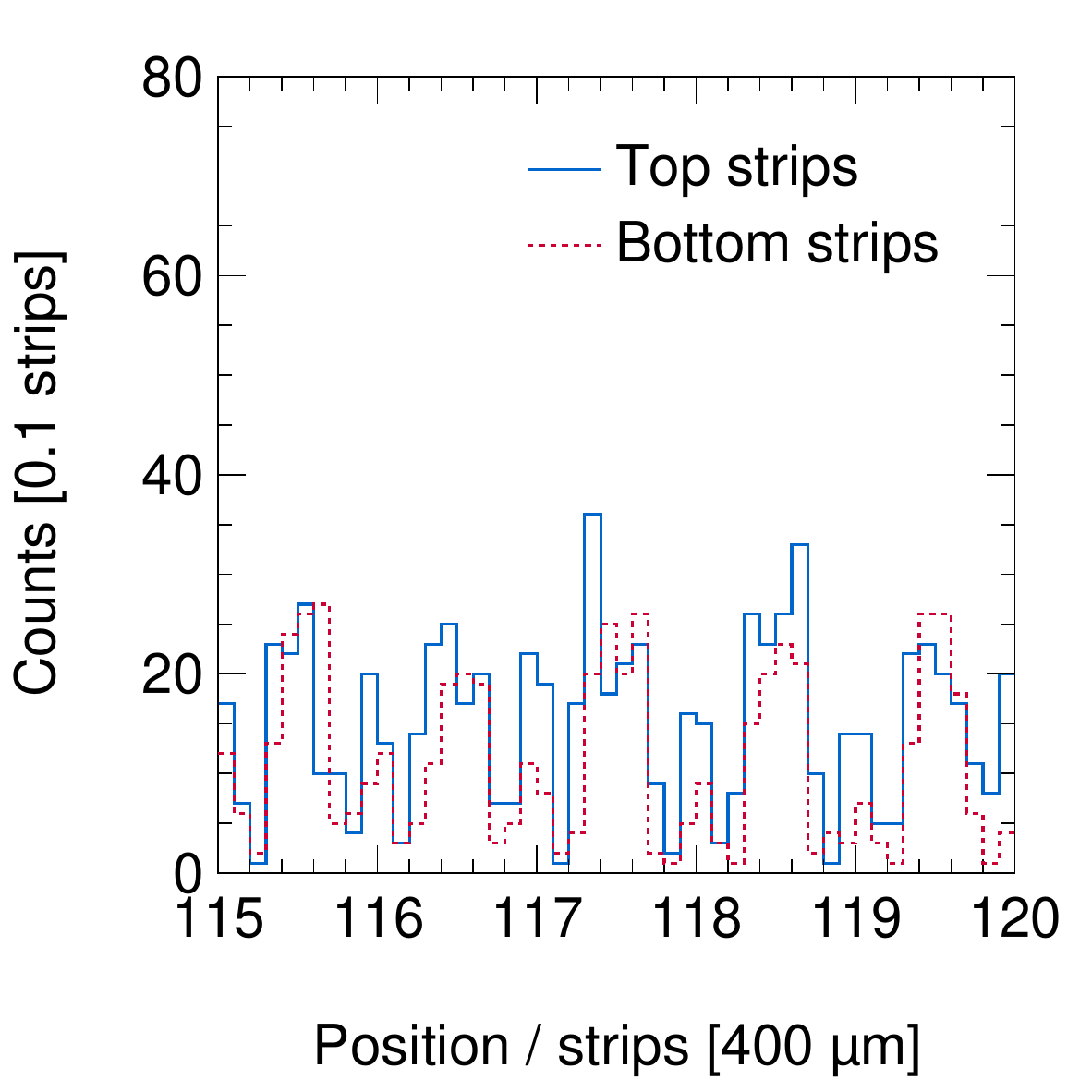}
        \caption{Low gain, approximately $\num{8000}$}
        \label{fig:readout-modulation-low-gain}
    \end{subfigure}
    \caption{Distribution of the reconstructed interaction points at high granularity for the MicroMegas detector at high detector gains (a) and low detector gains (b).}
    \label{fig:readout-modulation}
\end{figure}

Another observation is that the spatial resolution for the \textmu{}RWELL is almost the same for the two readout planes, as it is expected from equal charge sharing, while for the MicroMegas the behaviour between the top and bottom strips deviates.
This was found to be the result of the so-called `readout modulation', as illustrated by Fig.~\ref{fig:readout-modulation}.
It shows the distribution of the reconstructed interaction points with high granularity.
This makes a peak structure visible which originates from the readout pattern modulated into the position distribution that is expected to be uniform due to the detector's uniform irradiation.
Due to this modulation effect by the discrete readout structure in combination with a threshold-based zero-suppressed readout electronics, the interaction points are more likely to be reconstructed to the central strip for odd-strip-count clusters and in-between the two central strips for even-strip-count clusters.
For a more detailed description of this effect, it is referred to \cite{article_x-ray}, while it should be noted that the effect has already been observed with Multi-Wire Proportional Chambers \cite{mwpc}.
This behaviour affects the accuracy of the position determination, with a stronger effect leading to a worse spatial resolution.
This is what can be conducted from Fig.~\ref{fig:readout-modulation}, where the modulation effect is stronger for the top strips than for the bottom strips.
It seems to be an intrinsic behaviour of the detector as it is also observed in laboratory measurements using an \textsuperscript{55}Fe source.
A possible explanation that was found for this behaviour is the way the signal induction.
With the top strips being much thinner than the bottom strips, the distribution of the induced signal charge changes compared to the bottom strips, with more charge being in the cluster's central strip for the top strips and less charge being acquired in the outer strips.

\subsection{Improving the spatial resolution}

While so far, the results have been obtained with COG to calculate the position, previous studies \cite{article_x-ray,article_vmm-mpgd} showed that a different weighting of the charge in the COG formula
\begin{align}
    x' = \frac{\sum_i Q_i^n x_i}{\sum_i Q_i^n}
\end{align}
with $i$ the index of the readout strip with a signal above THL and $n$ the weighting factor reduces the modulation effect and thus improves the spatial resolution.
This is because of low amplitude signals in the tails of the charge distribution that go above the THL on one side of the charge distribution but not on the other one.
Thus, a certain fraction of charge information to reconstruct the position is lost and the reconstruction of the interaction point is forced towards the signal above the THL.
To reduce the weight of the charge distribution's tails and thus mitigate the bias in the calculation of the interaction point, $n > 1$ can be selected.
In previous studies with triple-GEM detectors \cite{article_vmm-mpgd}, $n = 2$ was used due to its simplicity and its proximity to the optimal solution obtained from simulation studies \cite{thesis_heikki,thesis_djunes}.
Thus, it was investigated, if this $Q^2$ weighting can also be used to improve the spatial resolution of \textmu{}RWELL and resistive MicroMegas detectors.

In addition, an orthogonal approach was investigated, making use of a hardware feature of the VMM3a front-end ASIC, its Neighbouring-Logic (NL).
By default, the neighbouring-logic is turned off, but when enabled, the NL triggers the acquisition of an induced signal with an amplitude below the THL, if the neighbouring channel has a signal that surpasses the THL.
This allows to obtain more charge information and thus to improve the reconstructed position.

\begin{figure}[t!]
    \centering
    \begin{subfigure}{0.4\columnwidth}
        \centering
        \includegraphics[width = \columnwidth]{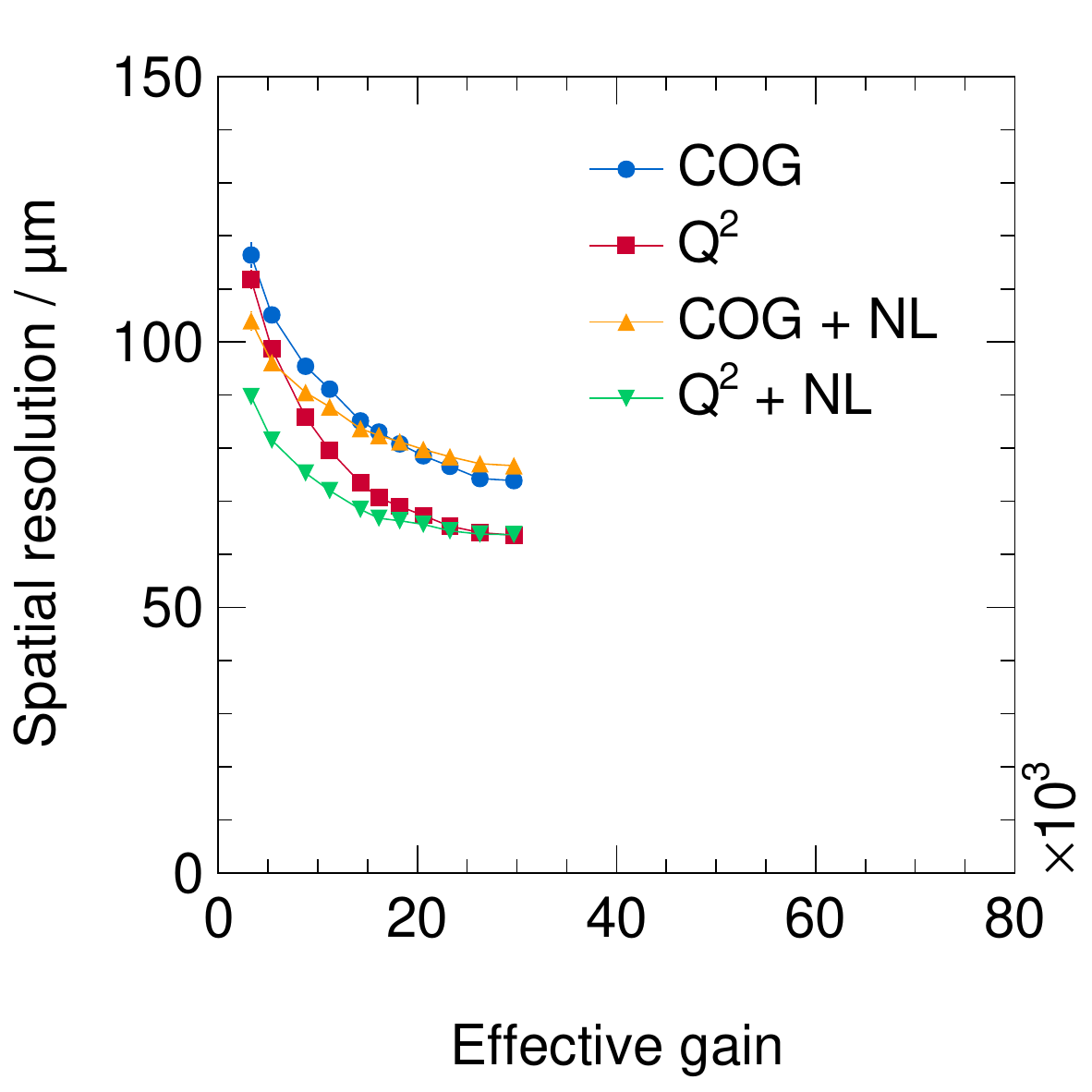}
        \caption{\textmu{}RWELL}
        \label{fig:spatial-resolution-gain-improved-U}
    \end{subfigure}
    \hspace{0.1\columnwidth}
    \begin{subfigure}{0.4\columnwidth}
        \centering
        \includegraphics[width = \columnwidth]{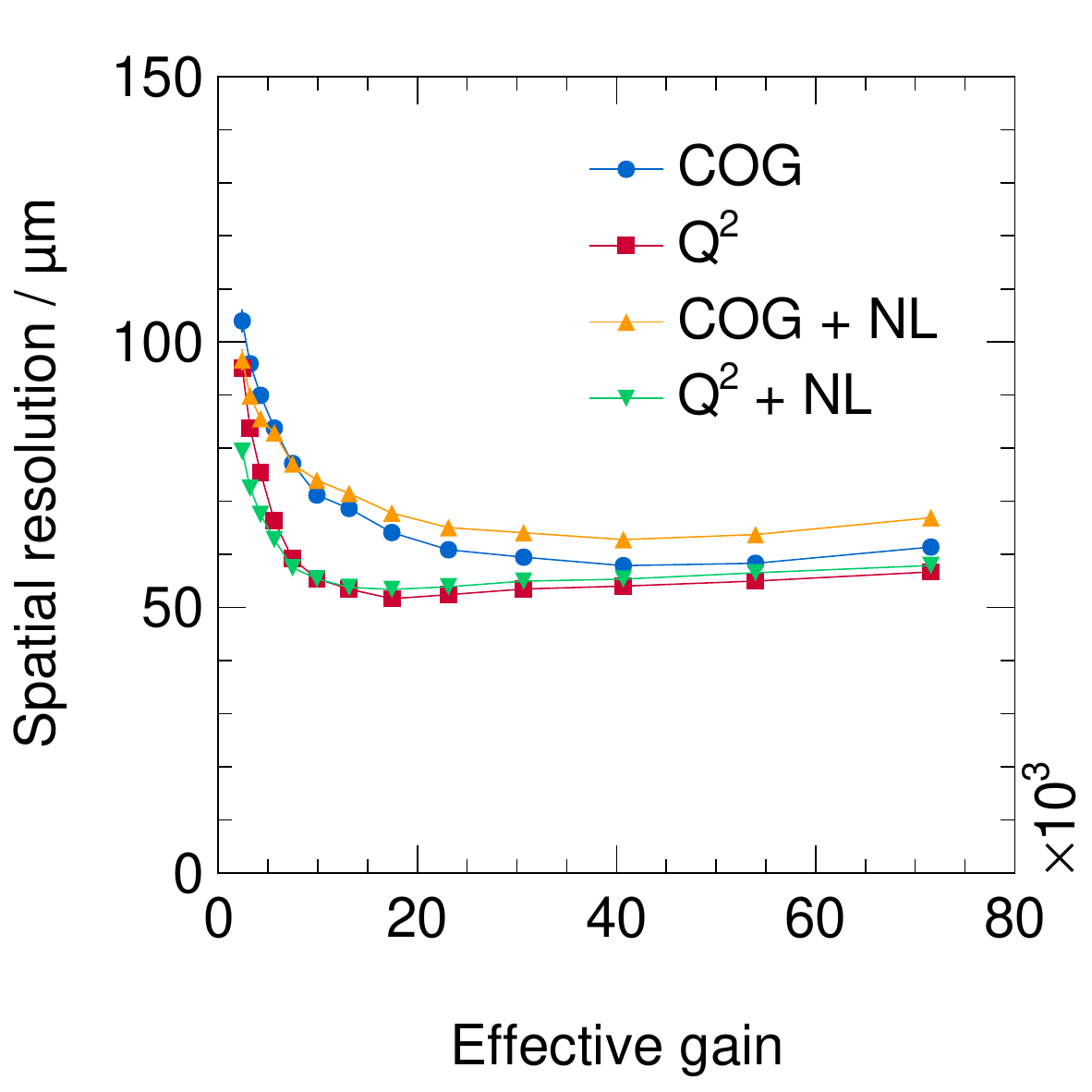}
        \caption{MicroMegas}
        \label{fig:spatial-resolution-gain-improved-MM}
    \end{subfigure}
    \caption{Dependence of the spatial resolution on the effective detector gain for the two different methods to determine the interaction point, with and without NL enabled.
    Here only the results of the bottom strips are shown, but the observed behaviour is the same for the top strops.}
    \label{fig:spatial-resolution-gain-improved}
\end{figure}

The results of these two methods, including their combination, are shown in comparison with the COG reconstruction in Fig.~\ref{fig:spatial-resolution-gain-improved}.
It can be seen that the $Q^2$ weighting has a large impact and significantly improves the spatial resolution, in both detector cases.
On the other hand, the effect of the NL is only visible at low gains, i.e. at low efficiencies and low signal-to-threshold ratios, where the relative amount of collected charge on the total cluster charge is large, with a large fraction of the induced signal charge being still below the THL.
At higher gains, this effect gets reduced, which is also reflected in the larger cluster size, i.e. on more channels a signal above THL was acquired.
Thus, the tails of the measured charge distribution contain less induced signal charge and the probability of acquiring electronics noise increases.
As a result, the spatial resolution decreases.
These two outcomes are exactly in line with what has been observed in previous studies \cite{article_vmm-mpgd,mythesis}.

\section{Time resolution}

As the last part of the characterisation studies, the time resolution of the two detectors is investigated.
For this, the interaction time of the particle measured in the DUTs is compared with the reference time measured with the scintillator/PMT/NIM-coincidence-unit combination, both acquired with VMM3a/SRS.
The reference time is provided as a constant amplitude signal --- due to the NIM output of the coincidence unit --- on a single channel of the VMM3a.
The measured interaction time in the DUTs is defined as
\begin{align}
    t' = \frac{1}{2}(t_\mathrm{top} + t_\mathrm{bottom}) \ .
\end{align}
The interaction timestamps for each plane, $t_\mathrm{top}$ and $t_\mathrm{bottom}$, correspond to the time of the signal within each cluster with the largest peak amplitude.

\begin{figure}[t!]
    \centering
    \begin{subfigure}{0.4\columnwidth}
        \centering
        \includegraphics[width = \columnwidth]{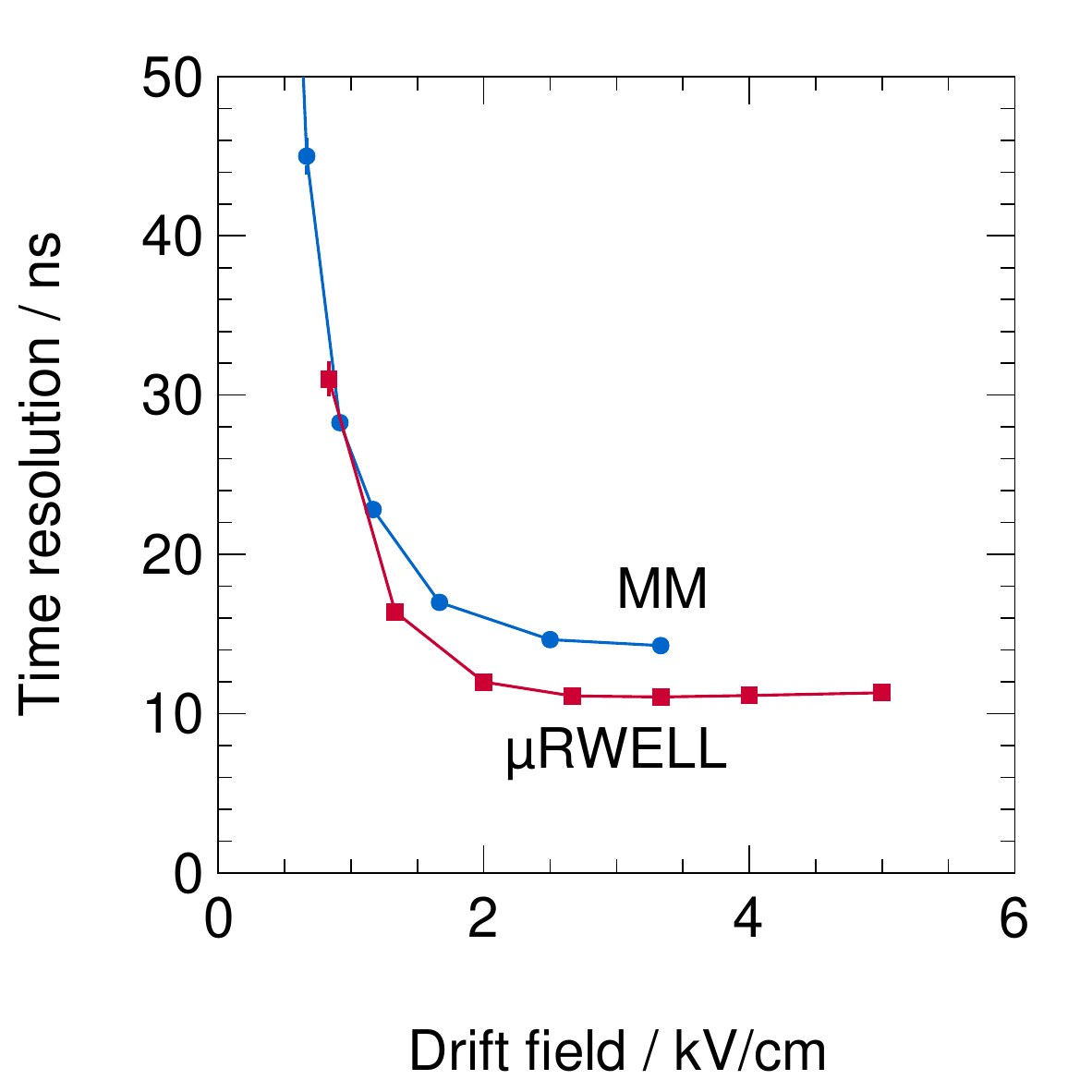}
        \caption{Field dependence}
        \label{fig:time-resolution-field}
    \end{subfigure}
    \hspace{0.1\columnwidth}
    \begin{subfigure}{0.4\columnwidth}
        \centering
        \includegraphics[width = \columnwidth]{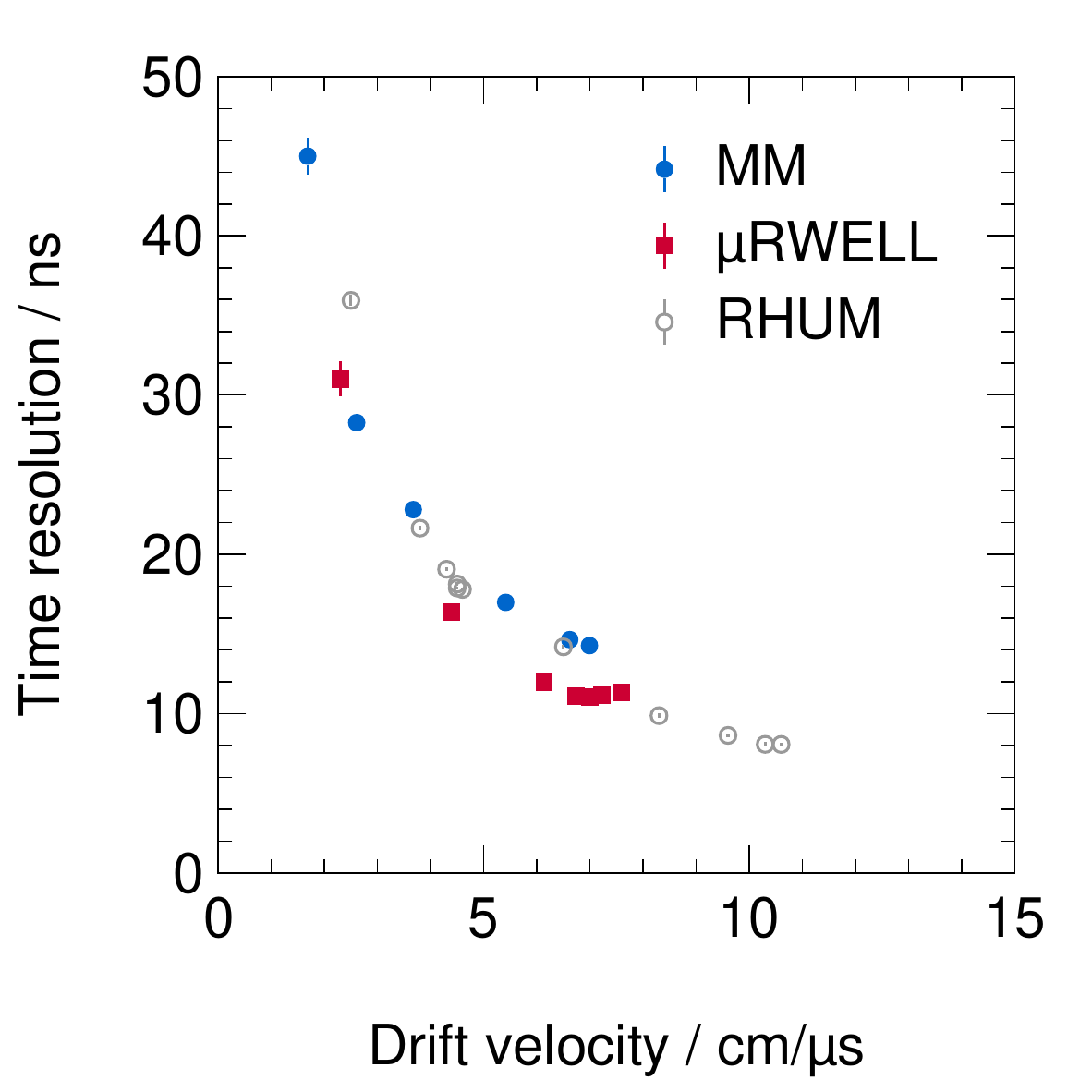}
        \caption{Velocity dependence}
        \label{fig:time-resolution-velocity}
    \end{subfigure}
    \caption{Measured time resolution for the two different detectors, depending on the electric drift field (a) and the electron drift velocity (b).
    For comparison purposes, also the results from measurements with a small-pad resistive MicroMegas detector from the RHUM project have been added \cite{rhum}.}
    \label{fig:time-resolution}
\end{figure}

With both timestamps found, their difference $\Delta t = t_\mathrm{ref} - t'$ is calculated.
Fitted to the resulting distribution is a single Gaussian function, with its width $\sigma_{\Delta t}$.
The final time resolution is then this width, from which the time resolution contributions of the VMM3a (around $\SI{2}{ns}$ at the used peaking time of $\SI{200}{ns}$) and the scintillator/PMT/NIM-logic (around $\SI{1.5}{ns}$) are quadratically subtracted.
The obtained time resolutions have been plotted against the electric drift field (Fig.~\ref{fig:time-resolution-field}) and the electron drift velocity (Fig.~\ref{fig:time-resolution-velocity}) which was calculated through Magboltz \cite{website_magboltz}. 
Added for reference, are data from measurements with a small-pad resistive MicroMegas detector from the RHUM project \cite{rhum}.
All data points show the same trend of an improved time resolution with increasing drift velocity.
Two aspects should be noted though in regard to the used gas mixtures.
The RHUM data have been obtained with a mixture of  Ar/CO\textsubscript{2}/iC\textsubscript{4}H\textsubscript{10} (93/5/2\,\%) for the low drift velocities and Ar/CF\textsubscript{4}/iC\textsubscript{4}H\textsubscript{10} (88/10/2\,\%) for the high drift velocities ($> \SI{6}{cm/\micro s}$).
In the case of the \textmu{}RWELL, which has been filled with Ar/CO\textsubscript{2} (70/30\,\%), the drift velocity does not change significantly at fields above $\SI{2.5}{kV/cm}$ --- a linear increase of around $\SI{1}{cm/\micro s}$ over a range of $\SI{2.5}{kV}$.
This explains the observed saturation behaviour of the time resolution.
In the case of the MicroMegas detector, this saturation behaviour could not be observed, because of the high-voltage power supply, which did not allow it to go to larger drift voltages.

In addition to the observation that the time resolution in the three detectors follows the same trend and that the results are compatible with each other, two other points are shown by the measurements.
Especially highlighted by the results from the resistive plane MicroMegas detector, it becomes clear that the working point for the best detector performance in terms of charge collection and spatial resolution (here at $\SI{0.9}{kV/cm}$) is not necessarily the working point for the best time resolution (here $>\SI{3}{kV/cm}$).
Secondly, due to the capabilities of VMM3a/SRS, both types of detectors, as well as the reference timing detectors, could all be read out by the same front-end electronics, with the corresponding data being contained in a single data stream.
This shows the versatility of the readout system, but also how it simplifies the data-taking and analysis process as it is not necessary to rely on additional high-precision timing electronics with an additional data stream that requires additional effort in the offline data analysis.

\section{Conclusion and outlook}

In this paper, the results from characterising two single-stage resistive plane MPGDs, a \textmu{}RWELL detector and a MicroMegas detector, with 2D strip readout have been presented.
Due to the capabilities of the new VMM3a/SRS readout electronics, it was possible to study simultaneously the charge behaviour, the spatial resolution and the time resolution of these two detectors.

One immediate observation was the difference in the amplitude of the induced signal measured by the front-end electronics between the two detectors, despite them being operated at the same gain.
This shows the importance of a good signal coupling between the anode and readout structure as otherwise, the detector efficiency can be still not sufficient although the detectors might be operated close to the breakdown voltages.
Both detectors showed good performance though.

The \textmu{}RWELL detector was investigated as a prototype detector for a future beam telescope of the DRD1 collaboration.
With time resolutions of $\SI{10}{ns}$, spatial resolutions of better than $\SI{70}{\micro m}$ in Ar/CO\textsubscript{2} (70/30\,\%) and equal charge sharing between the two readout strip planes, the detector fulfils the requirements to be used in the new telescope.
The MicroMegas detector was built with a 730 line-pairs-per-inch (LPI) mesh, corresponding to a mesh cell pitch of $\SI{35}{\micro m}$, in order to increase the operation stability.
Although the stability studies will be presented separately, the detector showed a good performance --- e.g.\ spatial resolutions of around $\SI{50}{\micro m}$ --- while being operated stably at gains of more than $\num{50000}$.
In addition, the results demonstrated that the optimal working point to achieve the best spatial resolution and charge collection, is not necessarily the best working point to achieve the best time resolution.

\acknowledgments
This work has been supported by the CERN EP R\&D Strategic Programme on Technologies for Future Experiments (\url{https://ep-rnd.web.cern.ch/}).

The authors would like to thank Jona Bortfeldt (LMU Munich) for providing the anamicom software (\url{https://gitlab.physik.uni-muenchen.de/Jonathan.Bortfeldt/anamicom}) to reconstruct the particle trajectories.

The authors would also like to thank Maria-Teresa Camerlingo (INFN Bari), Paolo Iengo (INFN Napoli), Mauro Iodice (INFN Roma Tre) and Marco Sessa (INFN Roma Tor Vergata) for the exchange and discussions on the MicroMegas results and in particular for providing the RHUM reference data shown in Fig.~\ref{fig:time-resolution}.


\begin{thebibliography}{99}

\bibitem{nsw}
T. Kawamoto et al. (The ATLAS Collaboration),
\emph{New Small Wheel Technical Design Report}, CERN-LHCC-2013-006, ATLAS-TDR-020 (2013).\\
URL: \url{https://cds.cern.ch/record/1552862}.

\bibitem{t2k}
D. Atti\'e et al.,
\emph{Characterization of resistive Micromegas detectors for the upgrade of the T2K Near Detector Time Projection Chambers},
\emph{Nucl. Instrum. Methods Phys. Res. A} {\bf 1025} (2022) 166109.\\
URL: \url{https://doi.org/10.1016/j.nima.2021.166109}.

\bibitem{rhum}
M. Iodice et al.,
\emph{Towards large size pixelized Micromegas for operation beyond 1 MHz/cm\textsuperscript{2}},
\emph{J. Instrum.} {\bf 18} (2023) C06029.

\bibitem{urwell-lhcb}
G. Bencivenni et al.,
\emph{The micro-RWELL detector for the phase-2 upgrade of the LHCb muon system},
\emph{Nucl. Instrum. Methods Phys. Res. A} {\bf 1049} (2023) 168075.\\
URL: \url{https://doi.org/10.1016/j.nima.2023.168075}

\bibitem{micromegas}
Y. Giomataris et al.,
\emph{MICROMEGAS: a high-granularity position-sensitive gaseous detector for high particle-flux environments},
\emph{Nucl. Instrum. Methods Phys. Res. A} {\bf 376} (1996) 29--35.\\
URL: \url{https://doi.org/10.1016/0168-9002(96)00175-1}

\bibitem{urwell}
G. Bencivenni et al.,
\emph{The micro-Resistive WELL detector: a compact spark-protected single amplification-stage MPGD},
\emph{J. Instrum.} {\bf 10} (2015) P02008.\\
URL: \url{https://doi.org/10.1088/1748-0221/10/02/P02008}

\bibitem{drd1-website}
DRD1 R\&D Collaboration,
\emph{Development of Gaseous Detectors Technologies}.\\
URL: \url{https://drd1.web.cern.ch/}

\bibitem{drd1-proposal}
A. Colaleo et al.,
\emph{DRD1 Extended R\&D Proposal}, CERN-DRDC-2024-003, DRDC-P-DRD1 (2024).\\
URL: \url{https://cds.cern.ch/record/2885937}.

\bibitem{website_rd51}
RD51 Collaboration,
\emph{Development of Micro-Pattern Gas Detectors Technologies}.\\
URL: \url{https://rd51-public.web.cern.ch/}

\bibitem{article_rd51}
M. Titov, L. Ropelewski,
\emph{Micro-Pattern Gaseous Detector Technologies and RD51 Collaboration},
\emph{Mod. Phys. Lett. A} {\bf 28} (2013) 1340022.\\
URL: \url{https://doi.org/10.1142/S0217732313400221}

\bibitem{article_vmm-mpgd}
L. Scharenberg et al.,
\emph{Performance of the new RD51 VMM3a/SRS beam telescope --- studying MPGDs simultaneously in energy, space and time at high rates},
\emph{J. Instrum.} {\bf 18} (2023) C05017.\\
URL: \url{https://doi.org/10.1088/1748-0221/18/05/C05017}

\bibitem{yi}
Y. Zhou et al.,
\emph{Fabrication and performance of a \textmu{}RWELL detector with Diamond-Like Carbon resistive electrode and two-dimensional readout},
\emph{Nucl. Instrum. Methods Phys. Res. A} {\bf 927} (2019) 31--36.\\
URL: \url{https://doi.org/10.1016/j.nima.2019.01.036}

\bibitem{article_vmm-twepp}
L. Scharenberg et al.,
\emph{Development of a high-rate scalable readout system for gaseous detectors},
\emph{J. Instrum.} {\bf 17} (2022) C12014.\\
URL: \url{https://doi.org/10.1088/1748-0221/17/12/C12014}

\bibitem{compass-gem}
C. Altunbas et al.,
\emph{Construction, test and commissioning of the triple-GEM tracking detector for COMPASS},
\emph{Nucl. Instrum. Methods Phys. Res. A} {\bf 490} (2002) 177--203.\\
DOI: \url{https://doi.org/10.1016/S0168-9002(02)00910-5}

\bibitem{vmm}
G. de Geronimo et al.,
\emph{The VMM3a ASIC},
\emph{IEEE Trans. Nucl. Sci.} {\bf 69} (2022) 976.\\
DOI: \url{https://doi.org/10.1109/TNS.2022.3155818}

\bibitem{michael}
M. Lupberger et al.,
\emph{Implementation of the VMM ASIC in the Scalable Readout System},
\emph{Nucl. Instrum. Methods Phys. Res. A} {\bf 903} (2018) 91--98.\\
DOI: \url{https://doi.org/10.1016/j.nima.2018.06.046}

\bibitem{srs}
S. Martoiu et al.,
\emph{Development of the scalable readout system for micro-pattern gas detectors and other applications},
\emph{J. Instrum. } {\bf 8} (2013) C03015.\\
DOI: \url{https://doi.org/10.1088/1748-0221/8/03/C03015}

\bibitem{thesis_jona}
J. Bortfeldt,
\emph{Development of Floating-Strip Micromegas Detectors},
PhD Thesis, Ludwig-Maximilians-Universität München (2014).\\
DOI: \url{https://doi.org/10.5282/edoc.16972}

\bibitem{thesis_horvat}
S. Horvat,
\emph{Study of the Higgs Discovery Potential in the Process $\mathit{pp}\rightarrow\mathit{H}\rightarrow 4\mu$},
PhD Thesis, University of Zagreb (2005).\\
DOI: \url{https://cds.cern.ch/record/858509}

\bibitem{mythesis}
L. Scharenberg,
\emph{Next-Generation Electronics for the Read-Out of Micro-Pattern Gaseous Detectors},
PhD Thesis, Rheinische Friedrich-Wilhelms-Universität Bonn (2023).\\
DOI: \url{https://hdl.handle.net/20.500.11811/10776}

\bibitem{article_x-ray}
L. Scharenberg et al.,
\emph{X-ray imaging with gaseous detectors using the VMM3a and the SRS},
\emph{Nucl. Instrum. Methods Phys. Res. A} {\bf 1011} (2021) 165576.\\
DOI: \url{https://doi.org/10.1016/j.nima.2021.165576}

\bibitem{mwpc}
F. Piuz et al.,
\emph{Evaluation of systematic errors in the avalanche localization along the wire with cathode strips read-out MWPC},
\emph{Nucl. Instrum. Methods Phys. Res.} {\bf 196} (1982) 451--462.\\
DOI: \url{https://doi.org/10.1016/0029-554X(82)90113-6}

\bibitem{thesis_heikki}
H. Pulkkinen,
\emph{Basic properties of Gas Electron Multipliers and different methods to determine the cluster characteristics},
Student Report Tfy-3.5111, Aalto University (2013).

\bibitem{thesis_djunes}
D. Janssens,
\emph{Resistive Electrodes and Particle Detectors: Modeling and Measurements of Novel Detector Structures},
PhD Thesis, Vrije Universiteit Brussels (2024).

\bibitem{website_magboltz}
S. F. Biagi,
\emph{Magboltz - transport of electrons in gas mixtures}\\
URL: \url{https://magboltz.web.cern.ch/magboltz/}

\end{thebibliography}
\end{document}